\documentclass[a4paper,fleqn,usenatbib]{mnras}

\usepackage{newtxtext,newtxmath}
\usepackage{txfonts}
\usepackage[T1]{fontenc}
\usepackage{ae,aecompl}
\usepackage{graphicx}	% Including figure files
\usepackage{amsmath}	% Advanced maths commands
\usepackage{amssymb}	% Extra maths symbols

\title[Star formation in Galactic flows]{Star formation in Galactic flows}

\author[R. Smilgys \& I. A. Bonnell]{
Romas Smilgys,$^{1}$\thanks{E-mail: rs202@st-andrews.ac.uk}
Ian A. Bonnell$^{1}$
\\
$^{1}$Scottish Universities Physics Alliance (SUPA), School of Physics and Astronomy, 
University of St. Andrews, North Haugh, St. Andrews, Fife KY16 9SS, UK
}

\date{Accepted XXX. Received YYY; in original form ZZZ}

\pubyear{2015}

\begin{document}
\label{firstpage}
\pagerange{\pageref{firstpage}--\pageref{lastpage}}
\maketitle

\begin{abstract}

We investigate the triggering of star formation in clouds that form in Galactic scale flows as the ISM passes through spiral shocks. We use
the Lagrangian nature of SPH simulations to trace how the star forming gas is gathered into self-gravitating cores that collapse to form stars. 
Large scale flows that arise due to Galactic dynamics create shocks of order 30 km s$^{-1}$ that compress the gas and form dense clouds  $(n> $several $\times 10^2$ cm$^{-3}$) in which self-gravity becomes relevant. These large-scale flows are necessary for creating the dense physical conditions for gravitational collapse and star formation. Local gravitational collapse requires densities in excess of $n>10^3$ cm$^{-3}$ which occur on size scales of $\approx 1$ pc for low-mass star forming regions ($M<100 M_{\odot}$), and up to sizes approaching 10 pc for higher-mass regions ($M>10^3 M_{\odot}$). Star formation in the 250 pc region lasts throughout the 5 Myr timescale of the simulation with a star formation rate of $\approx 10^{-1} M_{\odot}$ yr$^{-1}$ kpc$^{-2}$. In the absence of feedback, The efficiency of the star formation per free-fall time  varies from our assumed 100 \% at our sink accretion radius to values of $< 10^{-3}$ at low densities.

%This is a simple template for authors to write new MNRAS papers.
%The abstract should briefly describe the aims, methods, and main results of the paper.
%It should be a single paragraph not more than 250 words (200 words for Letters).
%No references should appear in the abstract.
\end{abstract}


%%%%%%%%%%%%%%%%%%%%%%%%%%%%%%%%%%%%%%%%%%%%%%%%%%

%%%%%%%%%%%%%%%%% BODY OF PAPER %%%%%%%%%%%%%%%%%%

\section{Introduction}

\begin{figure*}
	\includegraphics[width=2\columnwidth]{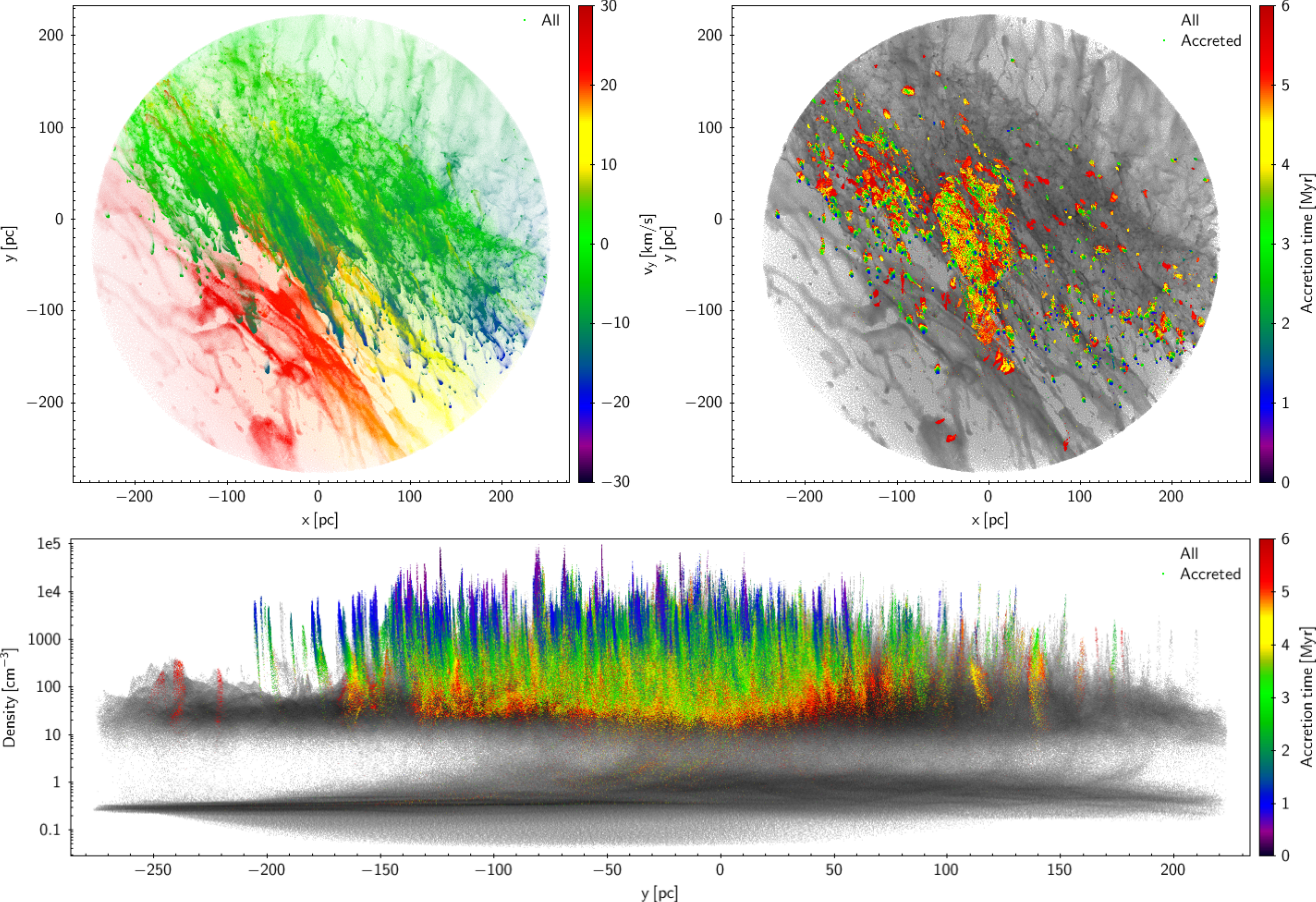}
	\caption{Initial conditions for accreting (coloured) and non-accreting (grey) particles. Colours in upper right panel show velocities in y direction and highlights the presence of 20-30 km/s spiral shock. Upper right panel and the low panel has accreted particles plotted on the top of all gas particles and colour coded by their accretion time. Accreted particles show that early star forming particles are located in highest density compact peaks, while late star forming particles are widespread in larger volumes.}
	\label{fig:iniCond}
\end{figure*}

Star formation is one of the most important processes in galactic evolution, transforming gas into stars and providing the visible output as well as the chemical and energetic feedback into the galaxy. Understanding how star formation is triggered is complex as it involves various physical processes that occur simultaneously, including Galactic scale and supernova driven flows that compress the interstellar gas, thermal physics, turbulence and the self-gravity of the gas that ultimately drives
the star formation process (\cite{2015ApJ...806...70L}).  Observationally, it is difficult to assess how star formation is triggered, and 
especially what are the initial conditions  that lead to  high-mass and clustered star formation (\cite{2011A&A...535A..76N}, \cite{2013A&A...555A.112P}, \cite{2014A&A...570A..15L}).

Significant work on star formation has centred on using idealised initial conditions to investigate the formation of stellar clusters and the origin of the initial mass function (\cite{2003MNRAS.339..577B}, \cite{2003MNRAS.343..413B}, \cite{2011MNRAS.410.2339B}, \cite{2011ApJ...740...74K}). \cite{2011MNRAS.410.2339B} investigated the evolution of an elongated cloud in which the gravitational binding varied along the major axis. This work showed that the initial conditions can have a major impact in all the resultant star formation properties
including star formation rates, efficiencies, clustering and even the IMF. A major conclusion is that using self-consistent initial conditions for star formation is crucial to develop realistic models. 

One way of generating initial conditions is using Galactic scale flows through either convergent streams (e.g. \cite{2007ApJ...657..870V}, \cite{2009MNRAS.398.1082B}, \cite{2012MNRAS.424.2599C}), or through spiral shocks (\cite{2006MNRAS.365...37B}, \cite{2006MNRAS.371.1663D}, \cite{2012MNRAS.425.2157D}, \cite{2013MNRAS.430.1790B}, \cite{2014MNRAS.441.1628S}). In the spiral shock case, the interstellar gas rotates at some 20-30 km/s faster than the pattern speed of the spiral arms. Adjacent streamlines are forced to converge in the spiral arms resulting in supersonic shocks which compress the gas, forming molecular clouds with regions that can become self-gravitating. Thermal instabilities in the shock result in formation of dense, cold ($\sim$10 K) star forming conditions even when the ISM is initially warm with temperatures between 1000 to $10^4$ K (\cite{2013MNRAS.430.1790B}).
Furthermore, the triggering also induces turbulent internal motions as are ubiquitously seen in molecular clouds (\cite{2006MNRAS.365...37B}, \cite{2013MNRAS.430.1790B}, \cite{2015MNRAS.446..973F}).

Galactic scale simulations of star formation \cite{2006MNRAS.371.1663D}, \cite{2013MNRAS.432..653D} have been used to study the formation and lifetimes of giant molecular clouds (GMCs). They found that the clouds form due to interactions of smaller clouds and the interstellar medium. The lifetimes of these clouds were in the range of 4-25 Myr for the highest mass clouds ($>10^5 M_{\odot}$), consistent with the local crossing time. The resolution of these simulations (with particle masses of $\sim$300 $M_{\odot}$) was  not high enough to resolve details of the star forming regions, or local gravitational collapse, within the individual clouds.  It is this aspect we aim to address here in order to study what drives the star formation within GMCs.

Star formation rates (SFR) and star formation efficiencies (SFE) are key parameters to characterise star forming region in terms of how much gas is turned into stars (\cite{2012ARA&A..50..531K}). The differences between these two parameters are that SFR describes the current rate - how much mass is going into star, while SFE tells how efficient is star formation for a given region, typically measured per dynamical, or free-fall time. Observationally it is challenging to determine SFE due to proper mass measurements; averaging subsets with different physical properties, which can't be properly resolved. \cite{2005ApJ...630..250K} argue that SFE per free-fall time should be constant independent of gas density.  
Observationally,  however, SFE are typically found to increase with gas density (\cite{2010A&A...524A..18B}, \cite{2013ApJ...762..120P}). 
A dynamical formation mechanism of molecular clouds due to Galactic scale flows help relax the constraints imposed if self-gravity is the primary formation mechanism (\cite{2011ApJ...740...74K}).
 Additional physical processes, such as feedback and magnetic fields can also act to delay star formation, or decrease the overall SFE. Collapse models including magnetic fields and turbulence find a significant reduction in SFRs in the range of 3-10
(\cite{2009MNRAS.398...33P}, \cite{2005ApJ...630..250K}, \cite{2011ApJ...730...40P}, \cite{2012ApJ...761..156F}, \cite{2012MNRAS.424..377D}, \cite{2013ApJ...770..150H}, \cite{2014MNRAS.442..694D}, \cite{2015A&A...573A.112M}).

In this paper, we analyse how star formation occurs in the spiral shocks simulation reported earlier in \cite{2013MNRAS.430.1790B}. The initial conditions arise from a nested suite of simulations starting from a full galaxy simulation, a local non-self gravitating cloud simulation and finally by a self-gravity simulation in which $1.1\times 10^6 M_{\odot}$ of gas is followed for over 5 million years. In Section 2 we describe this simulation in more detail and layout our analysis methods.

\section{Methods}

\begin{figure*}
	\includegraphics[width=2\columnwidth]{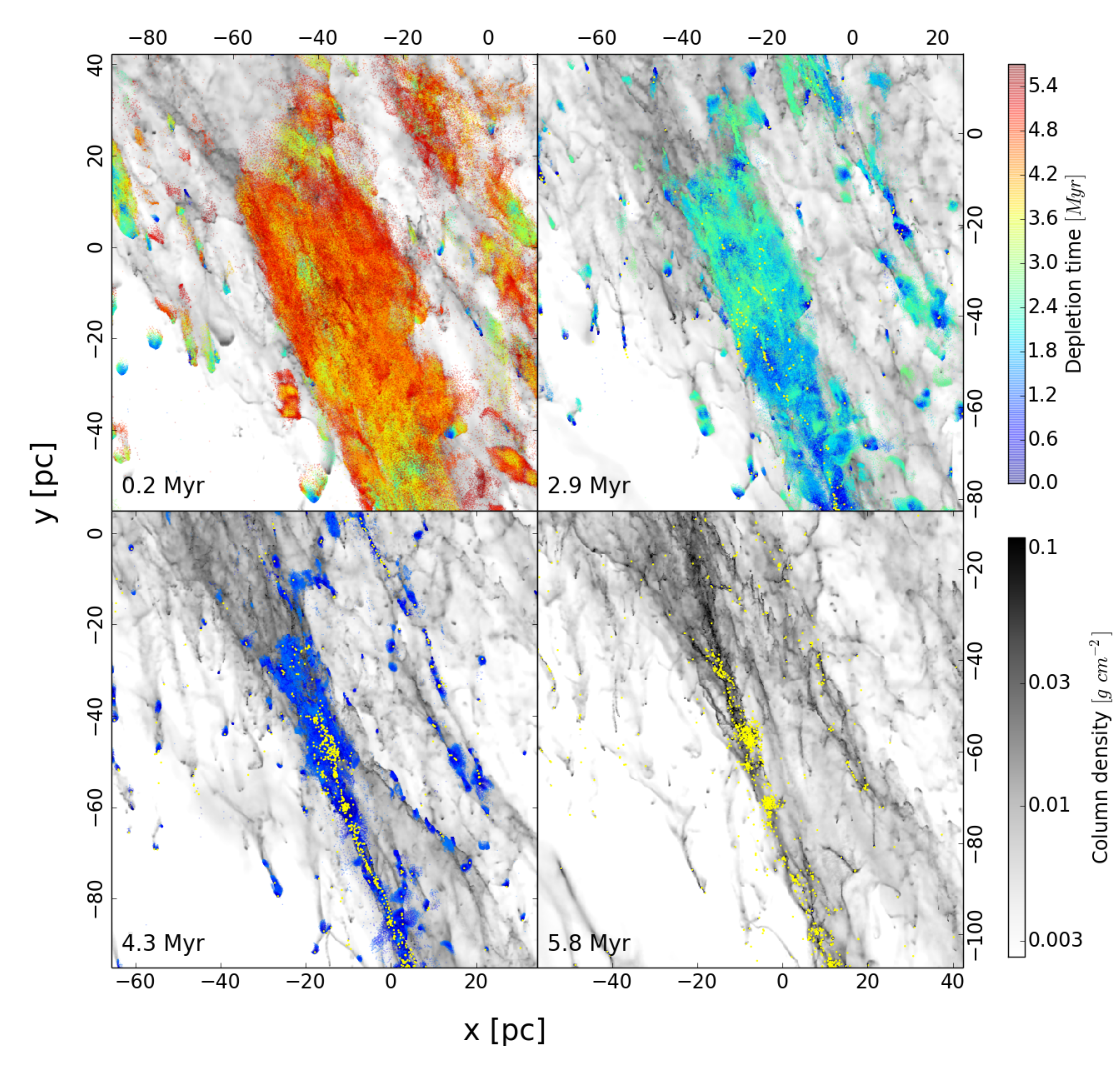}
	\caption{Evolutionary sequence of main star forming region. Grey colour represents column densities of all gas particles. Accreting particles are plotted on the top and colour coded by depletion time. Sink particles are shown as yellow dots. Four different panels show how clouds looks like at 0.2, 2.9, 4.3 and 5.8 Myr respectively. The presence of the main region collapse is visible over all times. Star formation take place in the densest parts of the region.}
	\label{fig:xyEvol}
\end{figure*}

\subsection{SPH simulations}

The work presented here is based on our analysis of a re-running of the {\it Gravity} simulation presented in \cite{2013MNRAS.430.1790B}. This simulation was constructed from a set of nested simulations starting from a full Galactic disc simulation over 350 Myr, with a 50 Myr high-resolution counterpart focusing on the formation of dense clouds in the spiral arms,
and a final simulation to follow star formation over $5.8$ Myr. The final
stage, which we analyse here,  included self-gravity and modelled a 250 pc region containing $1.9\times 10^6$ $M_{\odot}$ mass.
This simulation used $1.29\times 10^7$ SPH particles with 0.15 $M_{\odot}$ masses. Star formation is followed through the use of sink particles (\cite{1995MNRAS.277..362B}). A minimum mass for the sink particles corresponds to $\approx 70$ SPH particles representing one SPH kernel, or $\approx 11$ $M_{\odot}$ with a sink radius of $0.25$ pc to accrete bound, infalling gas particles while all particles penetrating within $0.1$pc were accreted. The sink particles therefore do not represent individual stars but rather a small cluster of stars or star forming region. Gravitational interactions between sinks were smoothed within $0.025$ pc.

The thermal physics was included as an adiabatic gas law with superimposed optically thin cooling rates
balanced against constant background UV heating (\cite{2002ApJ...564L..97K}, \cite{2007ApJ...657..870V}). This cooling function is an approximation based on the most important coolants at the relevant (low) densities including Lyman-alpha, CII and OI (\cite{2000ApJ...532..980K}). Use of a cooling function assumes that the gas is optically thin in the low density gas, with high density optically thick regions where gravitational collapse occurs being replaced by sink particles. 
The cooling rates provide a thermal instability once gas densities increases above $1$ ${\rm cm}^{-3}$ and results in   a multiphase ISM (see Fig. 1) where warm gas
at densities $<1$ cm$^{-3}$ and temperatures near $10^4$ K coexists with cooler gas down to temperatures approaching 10 K at densities in the range of $10$ to 1000 cm$^{-3}$ (\cite{2013MNRAS.430.1790B}).

Our use of a cooling curve is necessarily an approximation to the detailed cooling that would require an  explicit treatment of the chemical evolution in the gas (\cite{2012MNRAS.424.2599C}, \cite{2012MNRAS.421..116G}, \cite{2013MNRAS.432..626M}).
\cite{2013MNRAS.432..626M} conducted  detailed comparisons between cooling functions and chemical networks showing that overall the cooling function employed here does a reasonably good job at representing the thermodynamics. They reported fairly good agreement in the mass and volume fraction of the cold gas and properties of the dense clumps, with some differences in overall cloud morphology and post-shock velocities.
In the simulations reported here of Galactic-driven shocks of clumpy ISM, the velocities remain dominated by the large-scale flows, such that
any differences in the thermally-induced post-shock velocities are less important.

We note here that in addition to not including a treatment of the chemistry,  these simulations also do not include magnetic fields or feedback such as ionisation and supernovae. Feedback effects will mostly occur at later, post collapse stages in the star formation process. \cite{2015A&A...573A.112M} and \cite{2014MNRAS.442..694D} showed that feedback can act to decrease the star formation rate by up to 50 \%, but does not significantly alter the star formation process. Magnetic fields are more likely to affect the earlier stages of star formation by slowing down collapse (\cite{2008MNRAS.385.1820P}, \cite{2009MNRAS.398...33P}), and will be included in subsequent models. Our approach has been to include the physical processes individually to ensure that we understand their respective effects on the star formation process.

\subsection{Analysing the onset of star formation}

We make use of the Lagrangian nature of SPH, where gas elements are modelled by individual SPH particles, to follow the gas flows as the evolution proceeds towards gravitational collapse and star formation. We note the individual particles which form each sink particle, or are subsequently accreted by a given sink particle, and use these to trace the physical properties of the star forming gas. This is done for each sink, and at all times during the simulation including self-gravity. In order to characterise the evolution towards gravitational collapse, star formation and subsequent accretion in the simulation, we use the evolution of the half-mass radius of the gas which ultimately contributes to each sink. In addition, we follow the gas densities, the velocities, the kinetic, thermal and gravitational energies for each sink 
from the initial conditions until all its contributing gas is accreted. The masses over which these properties are calculated represent the final sink masses and are hence constant for each sink.

In order to compare the physical properties of star formation, we plot the relevant properties in terms of the sink age, measured relative to the initial formation of the sink particle. Negative sink ages measure the time before the sink is accreted and positive sink ages describe the accretion phase. We also note the time at which each gas particle is accreted onto a sink particle, the accretion time, and the location where this accretion occurred. 
We refer to the depletion time, which is the time remaining in the simulation to each particle before it is accreted. This is compared to predicted star formation, or compression  timescales for the free-fall collapse of the individual object, the larger scale region, or the compression time due to the large scale spiral shock.

\section{Flow dynamics inside spiral arm}

The first goal is to establish where in the simulation does the star formation come from. Figure \ref{fig:iniCond} shows the initial
conditions for the simulation in terms of the initial positions, velocities and densities of all gas particles. 
The top-left panel shows the full initial conditions, viewed  face on the galactic disc, and with particles colour coded by their velocity along the y axis.
The  large scale kinematics  are inherited from the global galactic potential and spiral arms.
Gas enters the spiral arm from the bottom left of the panel with positive y-velocities of 10-30 km/s, while gas already present in the arm has negative y-velocities as it rotates around the galaxy. This produces a convergent flow, and a shock of order 20-30 km/s, as the gas leaves the potential minimum of the arm. The shock induces turbulent motions in the gas (\cite{2013MNRAS.430.1790B}; \cite{2014NPGeo..21..587F}
), and forms high-density structured clouds where the star formation primarily occurs. 

The top-right panel in Figure \ref{fig:iniCond} highlights the location of the SPH particles that subsequently undergo star formation. These particles
are colour-coded by their accretion times, i.e. the time from the start of the simulation at which point they form, or are accreted onto, a sink particle.  Comparing the upper left and right panels shows that most of the star formation occurs in the central region where the colliding flows meet to compress the gas.  At the high gas densities produced by the shock,  subregions of the cloud  become gravitationally unstable and start to collapse with the first stars forming within  1-2 Myr of simulation time. 

We see early star forming gas (blue colour) in very compact clumps, while late star forming gas are distributed in much larger regions. This is in good agreement that early accreted gas are already close to their accretion points. Late star forming gas still has to make its way towards their sinks during the collapse, so they are still  far away from their sinks. Figure \ref{fig:iniCond} also
shows the same initial conditions plotted as the gas density versus the y-position and again colour-coded by accretion time. This figure highlights the multiphase aspect of the simulation with cold dense gas co-existing with warm, low density gas. The shocked region has already produced some very high-density clumps that  collapse fairly quickly on their free-fall times. Larger regions require further compression from the colliding flows and shock before becoming self-gravitating.

To better see how the region evolves with time, we plot a zoomed-in version of Figure \ref{fig:iniCond}'s upper-right panel for the central region at 4 different times in the evolution (Figure \ref{fig:xyEvol}). We show  column density maps representing all gas particles in grey. Accreted gas particles are colour coded by their depletion time, i.e. the actual time remaining till they will be accreted, while already formed sink particles are plotted as yellow dots. We see that the main region is collapsing in a parallel direction with the spiral shock front by contracting from $\sim20 - 30$ pc width down to several pc. This creates a long and thin high density ridge, extending several ten's of parsecs  from the upper left to lower right side of the diagram, at a time of  $\sim4$ Myr. Most intensive star formation is visible in the middle of this ridge. Finally at the end of simulation the ridge subdivides into a sequence of clusters, laying $\sim10$ pc away from each other.

\begin{figure}
	\includegraphics[width=\columnwidth]{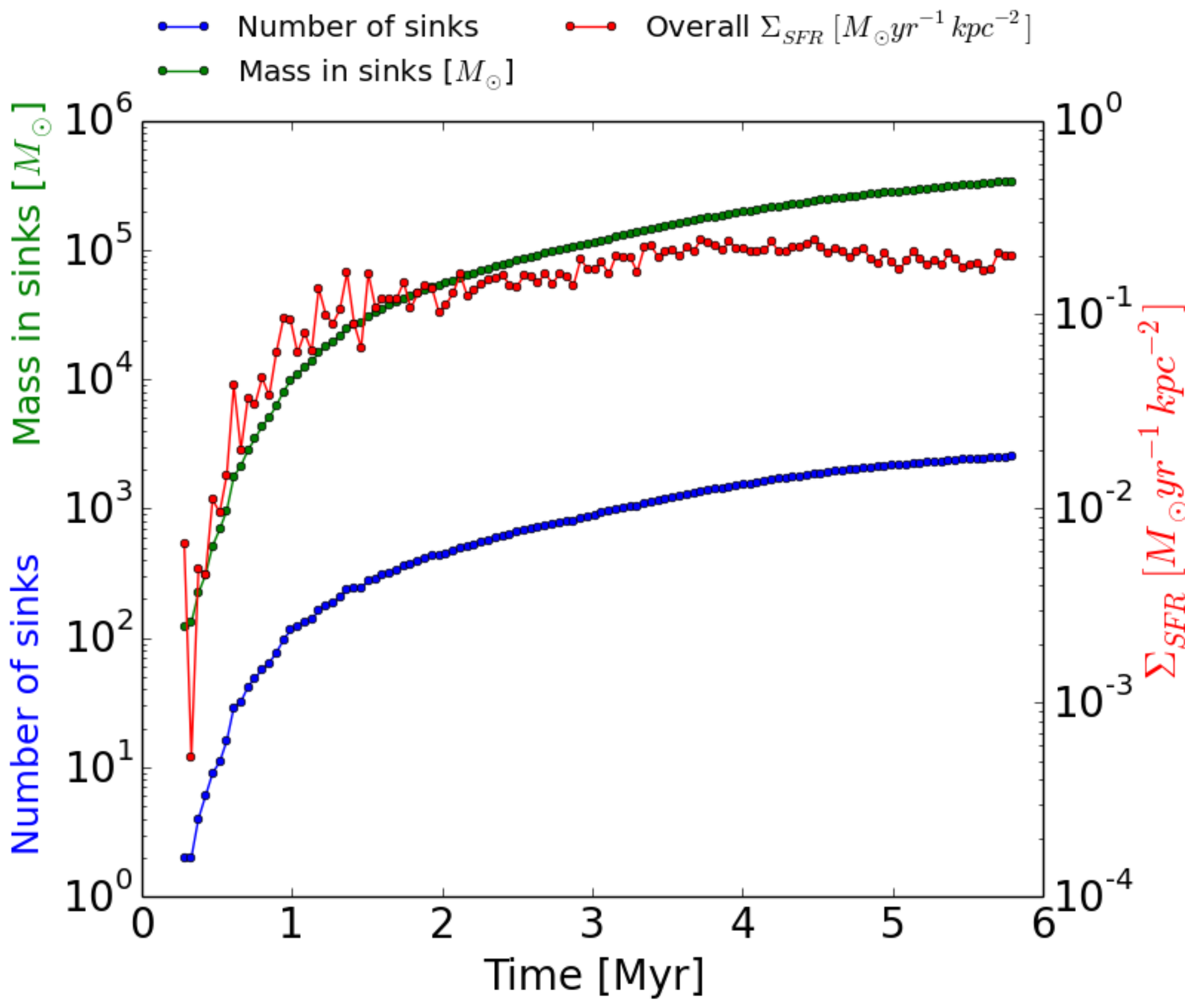}
	\caption{Simulation statistics in terms of number of sink particles, total mass in sinks and star formation rate as a function of time. Number of sinks and total mass in sinks increasing all the time, while $\Sigma_{SFR}$ grows very rapidly in the first Myr and at later times remains nearly constant at $\sim0.1-0.2$ $M_{\odot}yr^{-1}kpc^{-2}$.}
	\label{fig:stat}
\end{figure}

We show the statistics of the simulation as a function of time for sink particles in Figure \ref{fig:stat}. We see that the first sinks form  $\sim$0.5 Myr from the start of the simulation and at the end there are over 2000 sink particles with $\approx  4 \times 10^5$ M$_{\odot}$ in mass giving typical sink masses of $\approx 200$ $_{\odot}$ . The star formation increases rapidly from $0.5$ to $1.0$ Myr and saturates with a star formation rate of $\approx 0.1-0.2 $ $M_{\odot}yr^{-1}kpc^{-2}$. This star formation rate corresponds to an overall surface density of gas of $\approx 10$ $M_{\odot}$ $pc^{-2}$, placing it above the typical Schmidt-Kennicutt star formation rate. This discrepancy was also found in \cite{2013MNRAS.430.1790B}  and is likely due to the lack of magnetic support and potentially feedback (\cite{2015A&A...573A.112M}).

\section{Physics of star forming regions}

\begin{figure}
	\includegraphics[width=\columnwidth]{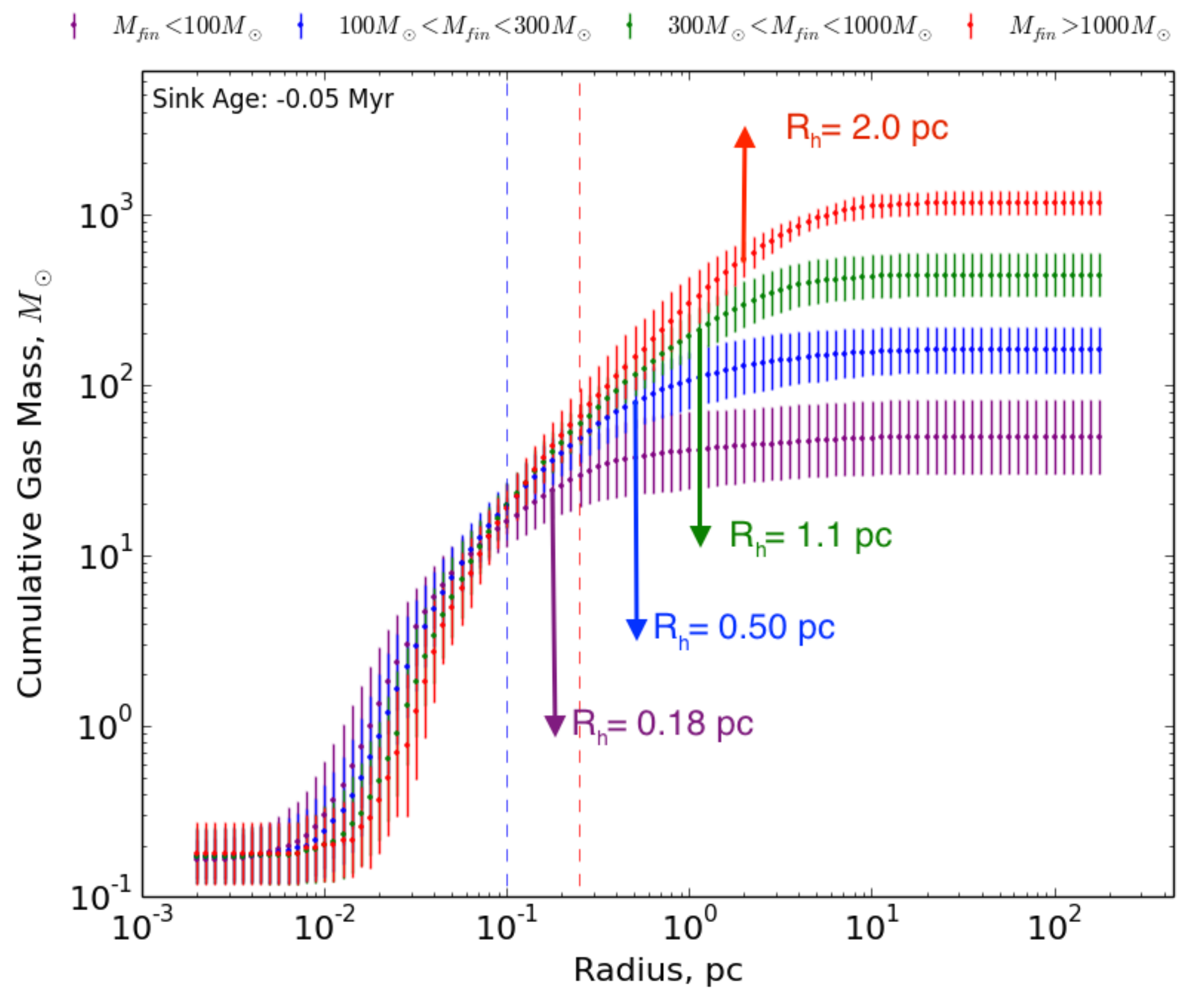}
	\caption{Cumulative mass distribution over radius for accreted gas particles at the time of sink formation. Blue vertical dashed line marks inner (0.1 pc), red - outer accretion radii (0.25 pc). Four colours indicates different final sink mass subsets. Low final mass sinks are forming with nearly all their final mass, while high final mass sinks collecting their high through accretion. Vertical arrows show mean half-mass radii for all four subsets.}
	\label{fig:cMassAccGas}
\end{figure}

\begin{figure}
	\includegraphics[width=\columnwidth]{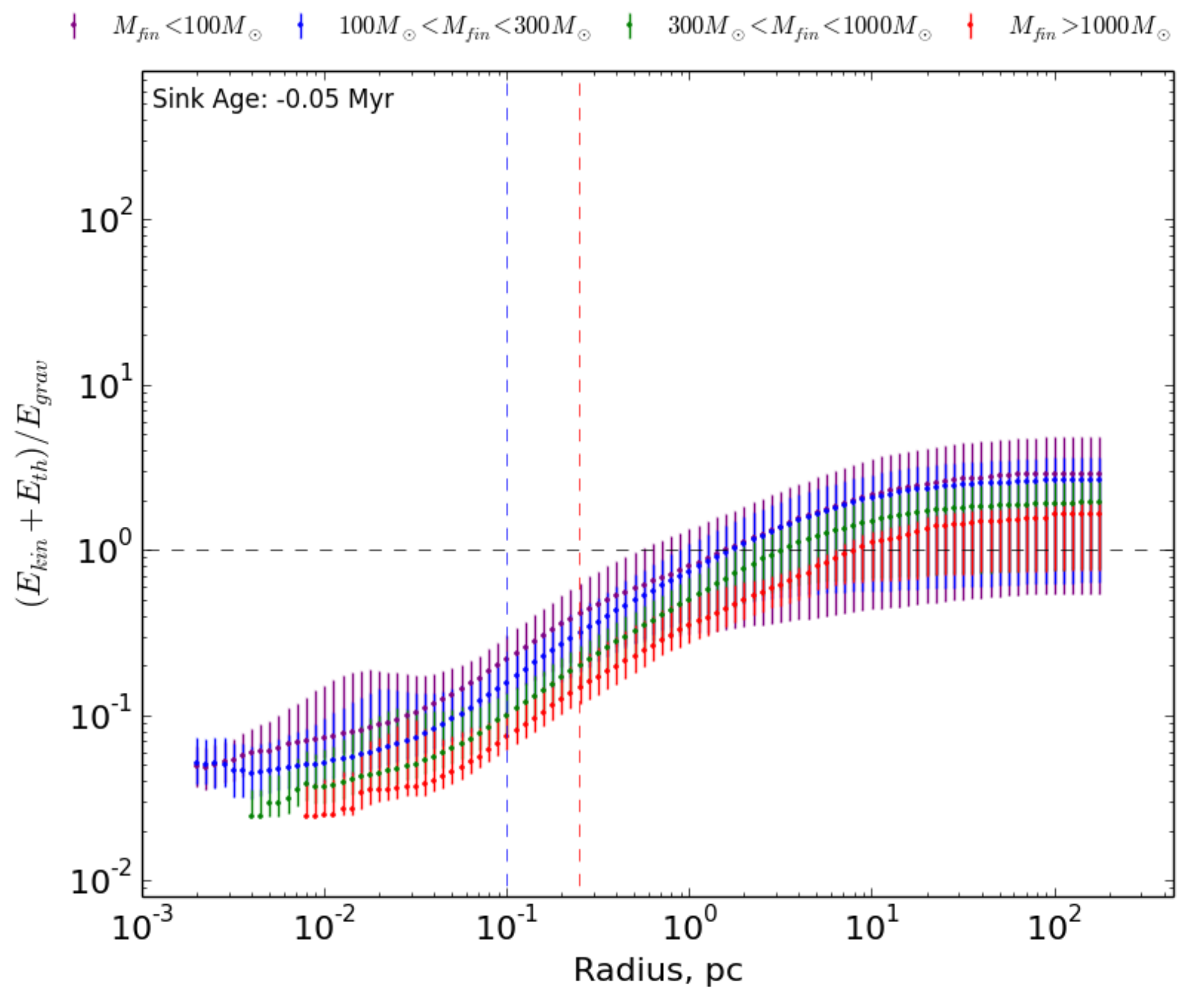}
	\caption{Binding ratios distribution over radius for accreted gas particles at the time of sink formation. Blue vertical dashed line marks inner (0.1 pc), red - outer accretion radii (0.25 pc). Black horizontal line (panel b) shows where sinks are gravitationally bound. Four colours indicates different final sink mass subsets. Larger final mass sinks are bound at larger scales than small final mass sinks.}
	\label{fig:cRatBind}
\end{figure}

\begin{figure}
	\includegraphics[width=\columnwidth]{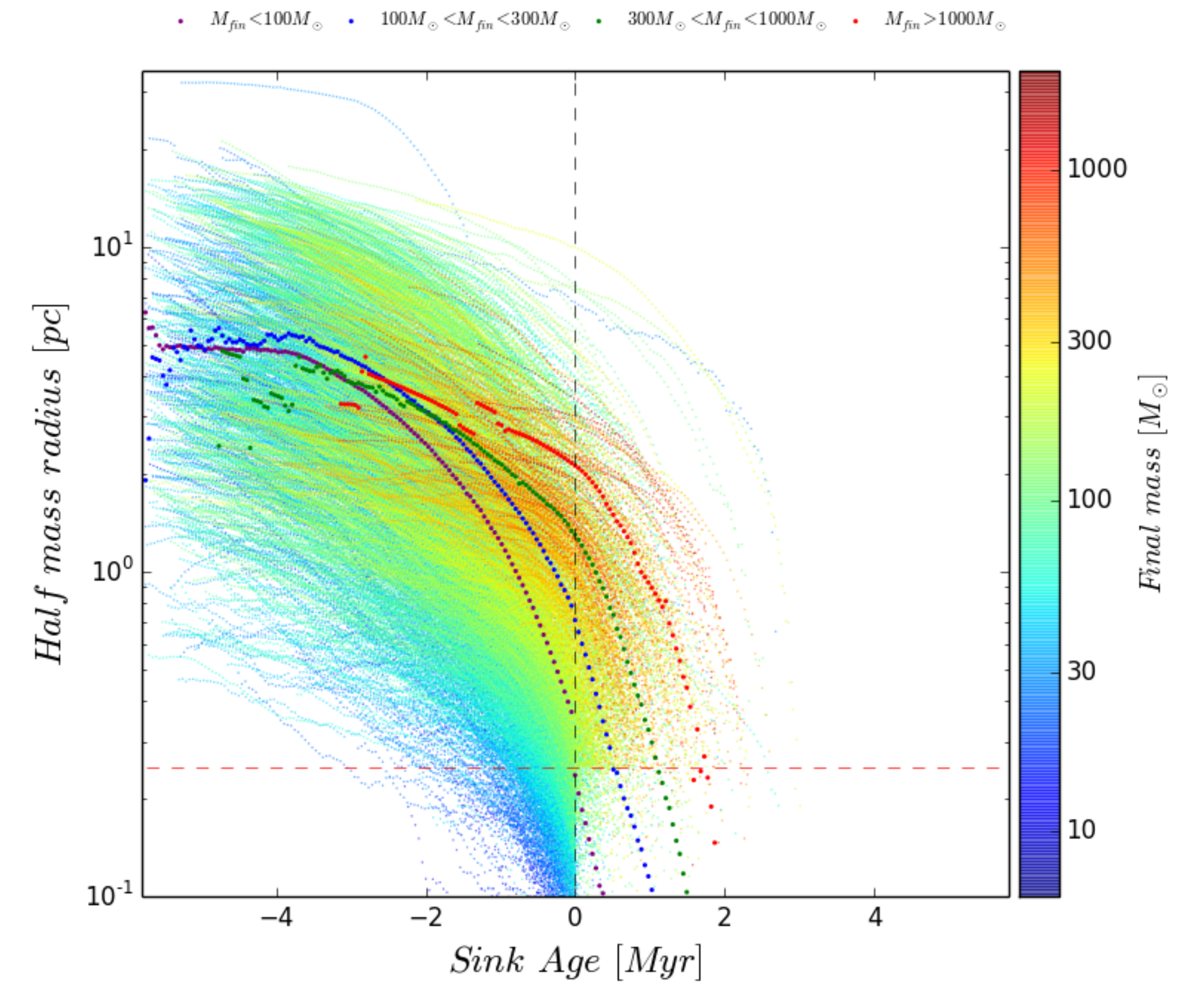}
	\caption{Evolution of sink half-mass radii (HMR) over sink ages. Tiny dots show sinks HMR tracks over sink ages and their colours represents final sink masses. Larger dots show paths of mean HMR evolution over sink ages for four final sink mass subsets. The diagram shows that HMR of low final mass sinks starts to decrease early all the way down to accretion radii. High final mass sinks sustains near the same HMR till the sink formation time and some of them starts to decrease their HMR only at 1-2 Myr after sink formation.}
	\label{fig:HMR}
\end{figure}

One of key questions which we need to understand  in terms of the physics of star formation is on what scale does gravity and hence star formation dominate over the background flows. We can use the Lagrangian nature of SPH to trace back the regions that contribute to the formation of individual sink particles, and measure their size distributions as well as their kinematic and gravitational energies.

\subsection{Accreted mass distribution at sink creation}

At the beginning of the simulation only gas particles are present. Sinks start to form after 0.5-1 Myr and can accrete throughout the simulation. We track the accreted gas particles for each sink and use these to analyse the physical properties of the sink at each point in the simulation.  This provides a constant number of particles to use for each sink and we use the term "final sink mass" to represent the sink mass at the end of simulation.  In order to aid comparison, we normalise the time sequence by the formation time for each sink.  

Before sink creation, we use the position of the particle that is subsequently turned into a sink, the proto-sink, as a reference centre for analysing the evolution to gravitational collapse and star formation. Figure \ref{fig:cMassAccGas} presents cumulative mass distributions with distance, averaged over four different final sink mass ranges,  for each sink at a time just before sink creation. This figure shows us how widely the mass is distributed relative to the sink accretion, or formation radius, of 0.25 pc.  In addition, the distributions vary significantly depending on the final sink mass.
Inside the sink accretion radius, the distributions are indistinguishable. All sinks
form with similar masses representing at least 70 SPH particles, as required numerically for sink creation within the simulation.
Outside the sink accretion radius, the mass distributions are very different. For low final-mass sinks ($m< 100$M$_{\odot}$),  Figure \ref{fig:cMassAccGas} shows that they do not accrete significantly from any envelope, with the majority of their final mass being located at sub-pc distances. 
Higher final-mass sinks
show increasing amounts of mass, located at larger distances from the newly created sink, that are yet to be accreted over longer timescales. The largest final-mass sinks, with masses in excess of 1000 $M_{\odot}$, have  about 100 times their formation mass in their envelopes, with the envelopes extending up to sizes of 10 pc.

\subsection{Binding ratios and half-mass radii}

To see at what scale  gravity becomes important, we calculated the kinetic, thermal and gravitational energies to check how bound individual particles are to their forming sink. Using the proto-sink as reference point, we use a direct summing of the gravitational energies and compare this to the kinetic and thermal energies. In Figure \ref{fig:cRatBind}, we plot the cumulative binding ratios, $(E_k+E_{th})/|E_g|$, for the sinks at the time of formation, again divided into four mass ranges. 
The binding ratios show that inner parts of the mass distributions for all forming sinks are bound to well outside the sink accretion radius. This is a necessary
condition for sink formation.
The difference lies in that low final-mass sinks are bound only to radii of $\sim1$ pc, whereas higher final-mass sinks are more deeply bound throughout, and are bound to greater distance of  $\sim5$ pc. The implication is that higher-mass sinks, or larger-mass star formatting regions, result from clumps that are more gravitationally bound, and where self-gravity dominates on larger sizescales.

We have seen above that, at the time of sink formation,  higher final-mass sinks still need to accrete significant amounts of their final mass which is located at relatively large distances. We  next calculated where this mass is located at earlier times, and how this region evolves in size throughout the simulation. To do this we determined the half-mass radius of the full final-mass distribution at each point in the simulation for each sink, or proto-sink. The half-mass radii are plotted as a function of sink ages in Figure \ref{fig:HMR}, for each individual sink and for mean values in four different mass ranges. We see a wide range of half-mass radii at early times extending from sub-pc to $>20$ pc. The
general evolution is that the half-mass radii decrease for all proto-sinks as they move closer to the time of sink formation. The
mean half-mass radii are $\sim5$ pc some 4 Myr before formation, apparently independent of final-mass. At later times, within 1 Myr of sink formation, the low and high final-mass sinks display different half-mass radii with the high final-mass sinks maintaining large half-mass radii of several pc while the  lower final-mass sinks halve mean half-mass radii of $<1$pc. 
Higher final-mass sinks maintain appreciable half-mass radii even after sink formation, with values of several pc up to 2 Myr after sink formation. This represents the region from which they subsequently accrete to gain their full masses.

\subsection{Driving of star formation}
\begin{figure}
	\includegraphics[width=\columnwidth]{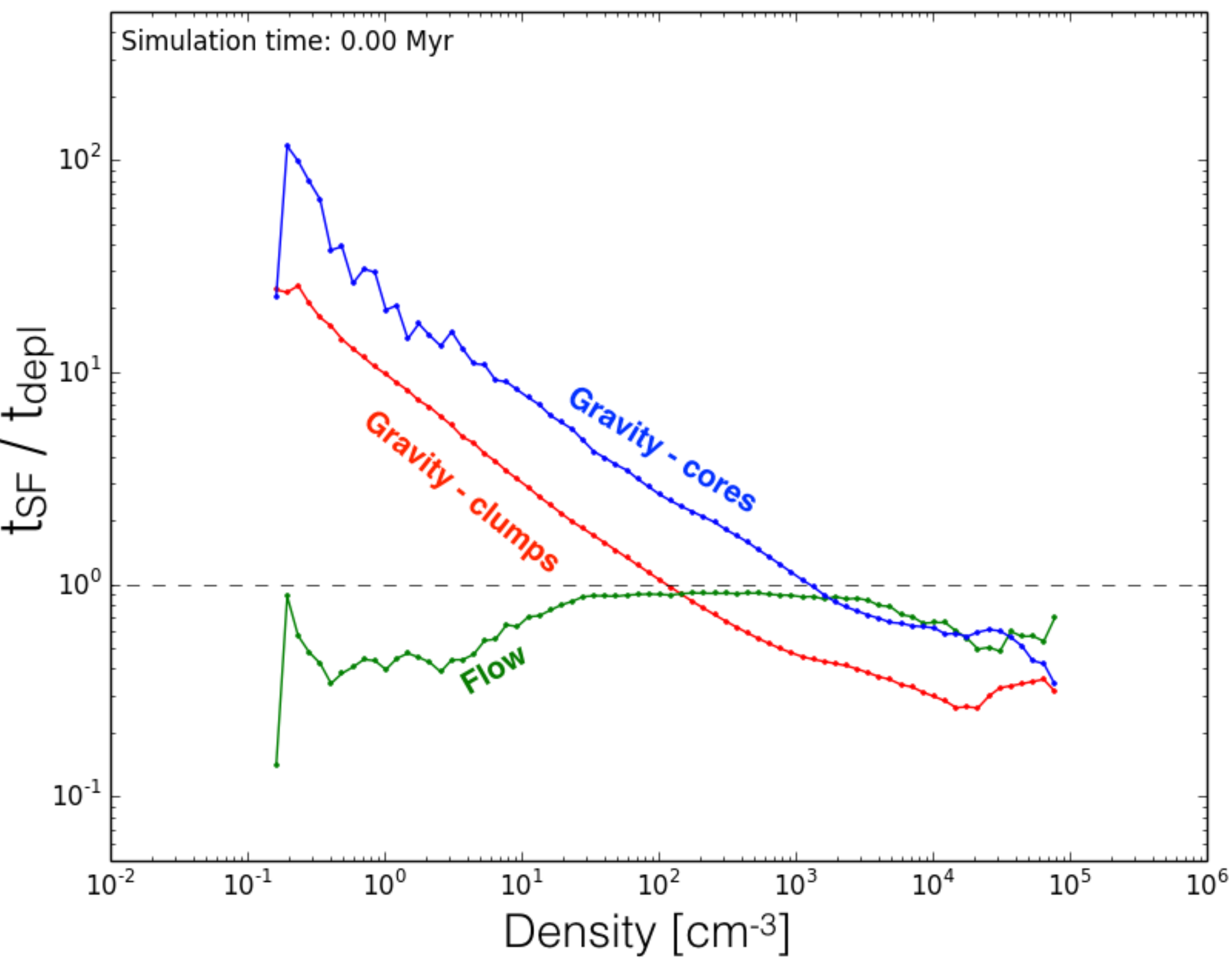}
	\caption{Ratio of star formation to depletion times vs SPH density histogram at the beginning of simulation. Colours represents different models for predicted star formation times.}
	\label{fig:tfftdepl}
\end{figure}

It is clear from Figure \ref{fig:iniCond} that star formation occurs mostly in the high density gas and that self-gravity, at least on small scales (Figure \ref{fig:cRatBind}) can drive the last phase of the star formation process. What is less clear is how the gas reaches the high densities required for gravity to force collapse. In order to assess this, we take all the gas particles that undergo star formation and calculate their {\sl depletion time}, the time until they are accreted, and compare this to three predictions for star formation timescales: (i) their {\sl free-fall time} or {\sl clump-collapse time} ($t_{clump} = \sqrt{3\pi/[32G\rho]}$) given their local density, (ii) their {\sl core-collapse time} ($t_{core} = \sqrt{{d_{sink}}^{3}/[GM_{sink}]}$) given the subset of their final mass which is interior to the particle, and (iii) the {\sl flow-time} given their local flow speed and distance ($d_{sink}$) towards the (proto-)sink ($t_{flow} = d_{sink}/v_{flow}$). This last is related to the large-scale flows due to the spiral shock.

All three processes gives independent prediction times for when particles should be accreted. In order to be able to physically explain the star formation observed in the simulation, we require that the predicted timescales $t_{pred}$, are less than the   timescale that is measured from the simulation ($t_{pred} \le t_{depl}$). The ratio of the three predicted timescales (the {\sl clump-collapse time} or free-fall time, the {\sl core-collapse time},  and the {\sl flow time}, or shock timescale)  are plotted in 
Figure \ref{fig:tfftdepl} binned by the  SPH particle densities. 
Blue dots represents {\sl core-collapse times} for the proto-sink where its enclosed mass is used. Red dots represent the free-fall time of the region {\sl clump-collapse times} as calculate by the gas particle's local gas density. It is a property of the environment and can inherently include effects of gas which does not contribute to the final-mass of the individual sink. The third model  (green dots) uses the flow velocity of each particle relative to the sink onto which it is to be accreted, and remaining distance to this (proto-) sink. We use only the radial velocity component towards the sink.  When the predicted times are greater than the observed depletion times, we can exclude that physical process from being responsible, at a given gas density, for driving the gas compression and ultimately star formation.

The diagram shows that at low densities (which represents large scales) the flow model is the only model which can explain the star formation, with $t_{flow}/t_{depl}\le 1$. Both gravity models at low densities  give large $t_{pred}/t_{depl}$ ratios. This is because gravitational collapse alone isn't enough to move accreted particles from these large distances sufficiently fast towards their sinks, regardless of the gas actually being bound or not at these scales. So galactic flows, which encompass large amount of radial velocities, are necessary in order to move the gas  from low to high density regions fast enough to explain the eventual star formation. At densities above a few $\times 10^2$ $cm^{-3}$, the gravity of the clump, given by the free-fall time of the region,  becomes short enough to give $t_{clump}/t_{depl}\le 1$. At this point we also see that the region's gravity model also crosses the flow model, which means that the gravity of the region is a significant process at these densities.. At  gas densities larger than  several $\times 10^3$ $cm^{-3}$ individual sinks' gravity becomes relevant with $t_{core}/t_{depl}\le 1$, although the individual core's self-gravity does not appear to dominate until much high densities are reached. The flow model continues to have low $t_{flow}/t_{depl}$ ratios but this is also expected in the gravitationally driven flows which dominate at these densities.

\subsection{Star Formation Efficiencies}

\begin{figure}
	\includegraphics[width=\columnwidth]{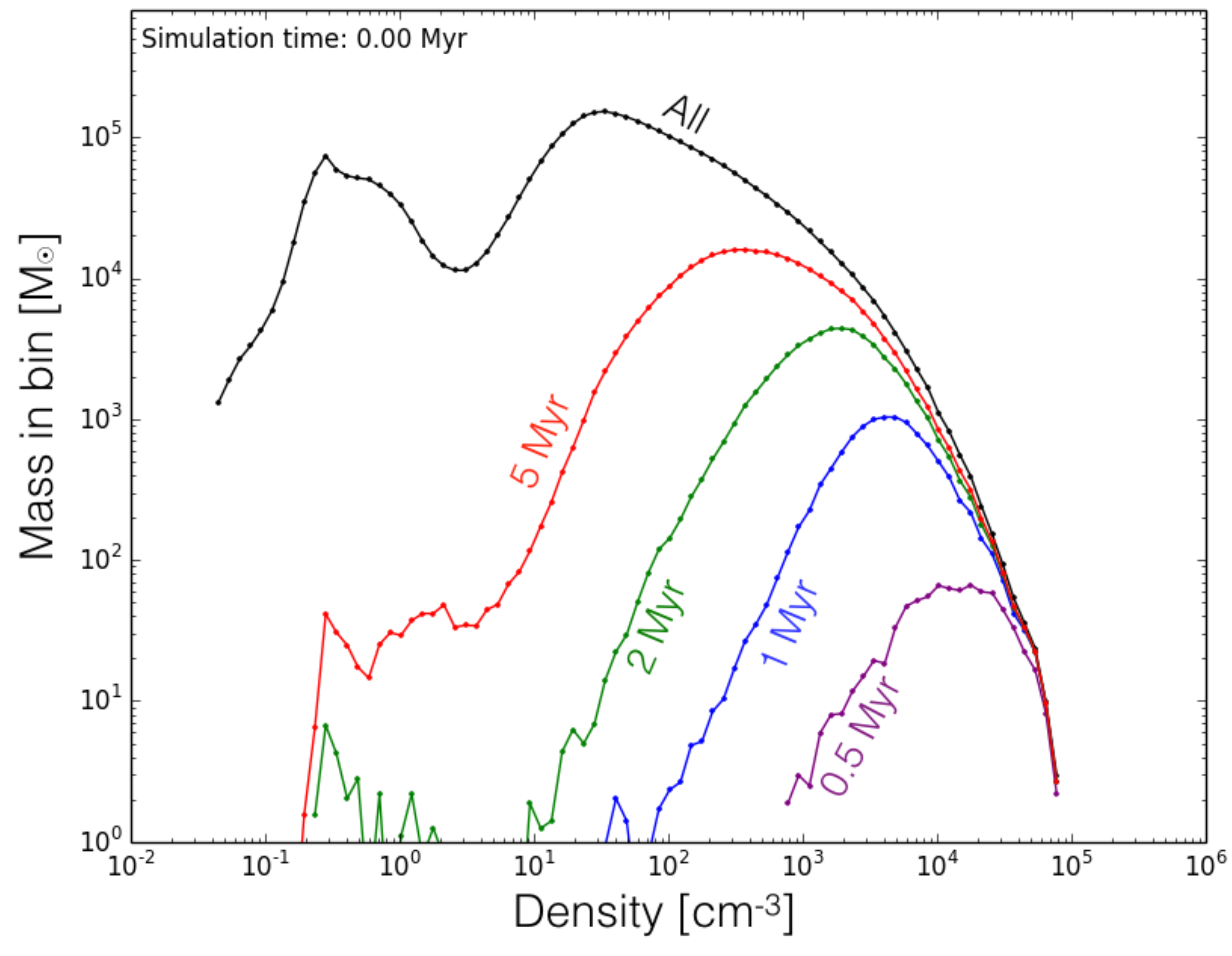}
	\caption{Mass distribution over SPH density for accreted and non-accreted gas at the beginning of simulation. Different colours shows mass distribution for particles being accreted only before given time.}
	\label{fig:Mass_Rho}
\end{figure}

We saw in the previous section that star formation is driven by the large scale flow from low gas densities until compression in the converging flow reaches sufficient densities for self-gravity to take over and local gravitational collapse ensues. A related issue is what fraction of the gas at a given density is involved in the star formation process. Figure \ref{fig:Mass_Rho} plots the  mass of gas as a function of its density at the start of the simulation when self gravity is turned on.  We see
the bimodal distribution of the gas as  a function of density, as evident in Figure~\ref{fig:iniCond}, where the cold dense gas is embedded in a warm, low density environment. Also plotted in Figure~\ref{fig:Mass_Rho} is the mass of gas in each density range that undergoes star formation within given times in the simulation. Gas at high densities ($n> 10^4$ cm$^{-3}$ ) undergoes star formation relatively quickly, with  of order 5 percent of gas at $10^4$ cm$^{-3}$ being accreted by sink particles within $0.5$ Myr and close to 20 percent by 1 Myr, reflecting free-fall time of $\approx 9\times 10^5$ years.  Gas at lower initial densities of $10^2$ cm$^{-3}$
sees only $\approx 0.1$ percent accreted within 2 Myr and $\approx 5$ percent within 5 Myr (free-fall time of $\approx 9 \times  10^6$ years).

\begin{figure}
	\includegraphics[width=\columnwidth]{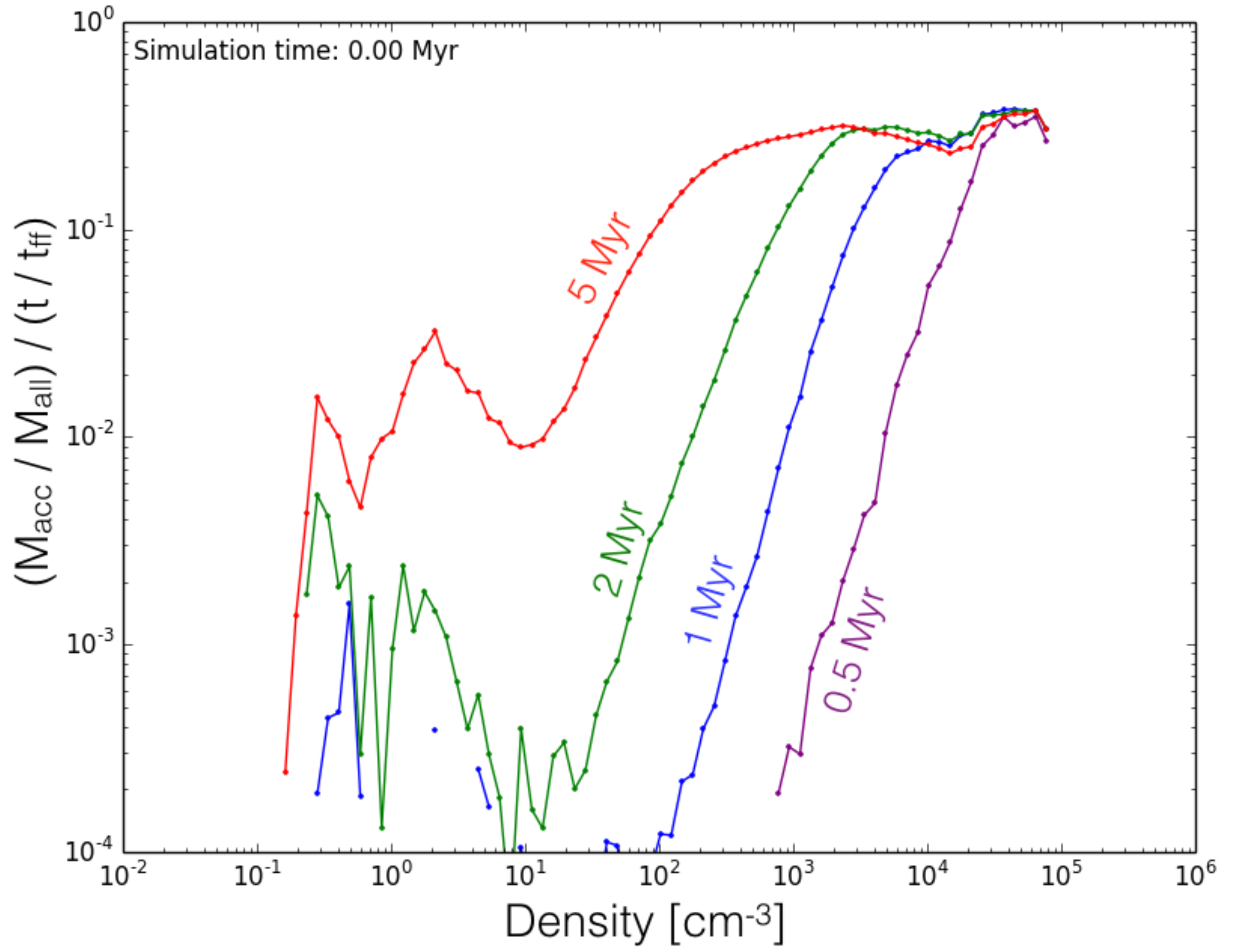}
	\caption{Star formation efficiencies (SFE) per free-fall time plotted as a function of the gas densities at the beginning of the simulation. Different colours shows mass distribution for particles being accreted only before given time. At large densities all SFE are high, however going towards lower densities SFE decreases depending to which timescales we are looking at. As gas are being accreted from lower density (larger distance) scales at longer times, there exist a point where feedback would halt continuous accretion from larger scales. If SNe feedback from first stars starts to play a role at $\sim$3-4 Myr, it is very likely that SFE lines would stop near the green line (2 Myr) and star formation efficiencies would drop significantly for densities bellow $10^2$ cm$^{-3}$, resulting in lower overall simulation SFE.}
	\label{fig:SFE}
\end{figure}

We see a similar trend with gas density for the star formation efficiency per free-fall time (Figure~\ref{fig:SFE}). 
We measure the star formation efficiency per free-fall time here  using our knowledge of the star formation events and timescales throughout the simulation. Thus, at a given gas density in the initial conditions of the simulation, we can measure the fraction of the gas that is accreted within a given time, relative to the free-fall time at that density. Although this measurement cannot be made observationally as it relies on the knowledge of the future evolution, it provides a more accurate measure of the actual efficiency of star formation. 
In Figure~\ref{fig:SFE}, we see at each time a generally increasing SFE from low to high densities, at which point it plateaus at roughly 30 percent per $t_{\rm ff}$. This overestimate is due to our assumption that 
accretion onto the sink particles is 100 per cent efficient, such that all gas particles that fall within the accretion radius of the sink contribute fully to the sink's mass-growth.  Regardless of this overestimation of the star formation efficiency, we see values are as low
as 2-3 \% at gas densities of $10^3$ cm$^{-3}$ after 1 Myr. 
These SFE estimations are comparable with 
\cite{2014A&A...570A..15L}, who  observed the high density regions in the ridge-like structure of W43-MM1 and obtains SFE as high as 3-11$\%$.
The SFE do increase with time, as one would expect in the absence of feedback or additional  support.
 Ensuring low star formation efficiencies over longer timescales requires additional physics such as feedback from ionising radiation or magnetic support which delays the collapse process.

\section{Conclusions}

We used SPH simulations to study the triggering of star formation from galactic scale flows. Our 1.9 million solar mass region of 250 pc produces approximately  400,000 solar masses in star formation over a 5.6 million year period. The large scale dynamics of the region clearly shows the compression of the gas in the shock region, and the subsequent star formation in the dense ridge of gas thus formed. The Lagrangian nature of SPH allowed us to trace back the star formation from the point of star formation, or subsequent accretion, to the earlier pre-collapse initial conditions. We found that at the point where star formation is initiated, our sink-formation time, lower mass regions are contained within $\approx 1$ pc while higher-mass regions extend for many 10's of pc. This is reflected in the size scale over which the region is bound. By investigating the timescales for gravitational collapse and for shock compression of the gas compared to the measured depletion timescale for an SPH particle to be accreted, we found that self-gravity is only relevant at high densities while the large scale compression due to the shock is responsible for driving star formation from low densities. Only a small fraction of gas initially at low densities is involved with star formation whereas the majority of the gas at higher densities is transformed into our sink particles. The efficiencies of star formation per free-fall time is correspondingly several 10 percent at densities of $10^5$ and decreases down to 0.1 percent at densities of several 100's cm$^{-3}$ within the first million years. From these results, we can conclude that in these simulations, and in the absence of magnetic fields, star formation is triggered by the large scale flows and that even at the point where star formation is initiated, the gas involved is spread over several, to several 10's of pc.

\section*{Acknowledgements}

RS and IAB acknowledges funding from the European Research Council for the FP7 ERC advanced grant project ECOGAL.
This work used the DiRAC Complexity system, operated by the University of Leicester IT Services, which forms part of the STFC DiRAC HPC Facility (www.dirac.ac.uk ). This equipment is funded by BIS National E-Infrastructure capital grant ST/K000373/1 and  STFC DiRAC Operations grant ST/K0003259/1. DiRAC is part of the National E-Infrastructure.

%%%%%%%%%%%%%%%%%%%%%%%%%%%%%%%%%%%%%%%%%%%%%%%%%%



\bibliographystyle{mnras}
\bibliography{mnrasSinks_arXiv} % if your bibtex file is called example.bib

\begin{thebibliography}{}
\makeatletter
\relax
\def\mn@urlcharsother{\let\do\@makeother \do\$\do\&\do\#\do\^\do\_\do\%\do\~}
\def\mn@doi{\begingroup\mn@urlcharsother \@ifnextchar [ {\mn@doi@}
  {\mn@doi@[]}}
\def\mn@doi@[#1]#2{\def\@tempa{#1}\ifx\@tempa\@empty \href
  {http://dx.doi.org/#2} {doi:#2}\else \href {http://dx.doi.org/#2} {#1}\fi
  \endgroup}
\def\mn@eprint#1#2{\mn@eprint@#1:#2::\@nil}
\def\mn@eprint@arXiv#1{\href {http://arxiv.org/abs/#1} {{\tt arXiv:#1}}}
\def\mn@eprint@dblp#1{\href {http://dblp.uni-trier.de/rec/bibtex/#1.xml}
  {dblp:#1}}
\def\mn@eprint@#1:#2:#3:#4\@nil{\def\@tempa {#1}\def\@tempb {#2}\def\@tempc
  {#3}\ifx \@tempc \@empty \let \@tempc \@tempb \let \@tempb \@tempa \fi \ifx
  \@tempb \@empty \def\@tempb {arXiv}\fi \@ifundefined
  {mn@eprint@\@tempb}{\@tempb:\@tempc}{\expandafter \expandafter \csname
  mn@eprint@\@tempb\endcsname \expandafter{\@tempc}}}

\bibitem[\protect\citeauthoryear{{Banerjee}, {V{\'a}zquez-Semadeni},
  {Hennebelle}  \& {Klessen}}{{Banerjee} et~al.}{2009}]{2009MNRAS.398.1082B}
{Banerjee} R.,  {V{\'a}zquez-Semadeni} E.,  {Hennebelle} P.,   {Klessen} R.~S.,
   2009, \mn@doi [\mnras] {10.1111/j.1365-2966.2009.15115.x}, \href
  {http://adsabs.harvard.edu/abs/2009MNRAS.398.1082B} {398, 1082}

\bibitem[\protect\citeauthoryear{{Bate}, {Bonnell}  \& {Price}}{{Bate}
  et~al.}{1995}]{1995MNRAS.277..362B}
{Bate} M.~R.,  {Bonnell} I.~A.,   {Price} N.~M.,  1995, \mnras, \href
  {http://adsabs.harvard.edu/abs/1995MNRAS.277..362B} {277, 362}

\bibitem[\protect\citeauthoryear{{Bate}, {Bonnell}  \& {Bromm}}{{Bate}
  et~al.}{2003}]{2003MNRAS.339..577B}
{Bate} M.~R.,  {Bonnell} I.~A.,   {Bromm} V.,  2003, \mn@doi [\mnras]
  {10.1046/j.1365-8711.2003.06210.x}, \href
  {http://adsabs.harvard.edu/abs/2003MNRAS.339..577B} {339, 577}

\bibitem[\protect\citeauthoryear{{Bonnell}, {Bate}  \& {Vine}}{{Bonnell}
  et~al.}{2003}]{2003MNRAS.343..413B}
{Bonnell} I.~A.,  {Bate} M.~R.,   {Vine} S.~G.,  2003, \mn@doi [\mnras]
  {10.1046/j.1365-8711.2003.06687.x}, \href
  {http://adsabs.harvard.edu/abs/2003MNRAS.343..413B} {343, 413}

\bibitem[\protect\citeauthoryear{{Bonnell}, {Dobbs}, {Robitaille}  \&
  {Pringle}}{{Bonnell} et~al.}{2006}]{2006MNRAS.365...37B}
{Bonnell} I.~A.,  {Dobbs} C.~L.,  {Robitaille} T.~P.,   {Pringle} J.~E.,  2006,
  \mn@doi [\mnras] {10.1111/j.1365-2966.2005.09657.x}, \href
  {http://adsabs.harvard.edu/abs/2006MNRAS.365...37B} {365, 37}

\bibitem[\protect\citeauthoryear{{Bonnell}, {Smith}, {Clark}  \&
  {Bate}}{{Bonnell} et~al.}{2011}]{2011MNRAS.410.2339B}
{Bonnell} I.~A.,  {Smith} R.~J.,  {Clark} P.~C.,   {Bate} M.~R.,  2011, \mn@doi
  [\mnras] {10.1111/j.1365-2966.2010.17603.x}, \href
  {http://adsabs.harvard.edu/abs/2011MNRAS.410.2339B} {410, 2339}

\bibitem[\protect\citeauthoryear{{Bonnell}, {Dobbs}  \& {Smith}}{{Bonnell}
  et~al.}{2013}]{2013MNRAS.430.1790B}
{Bonnell} I.~A.,  {Dobbs} C.~L.,   {Smith} R.~J.,  2013, \mn@doi [\mnras]
  {10.1093/mnras/stt004}, \href
  {http://adsabs.harvard.edu/abs/2013MNRAS.430.1790B} {430, 1790}

\bibitem[\protect\citeauthoryear{{Bontemps}, {Motte}, {Csengeri}  \&
  {Schneider}}{{Bontemps} et~al.}{2010}]{2010A&A...524A..18B}
{Bontemps} S.,  {Motte} F.,  {Csengeri} T.,   {Schneider} N.,  2010, \mn@doi
  [\aap] {10.1051/0004-6361/200913286}, \href
  {http://adsabs.harvard.edu/abs/2010A%26A...524A..18B} {524, A18}

\bibitem[\protect\citeauthoryear{{Clark}, {Glover}, {Klessen}  \&
  {Bonnell}}{{Clark} et~al.}{2012}]{2012MNRAS.424.2599C}
{Clark} P.~C.,  {Glover} S.~C.~O.,  {Klessen} R.~S.,   {Bonnell} I.~A.,  2012,
  \mn@doi [\mnras] {10.1111/j.1365-2966.2012.21259.x}, \href
  {http://adsabs.harvard.edu/abs/2012MNRAS.424.2599C} {424, 2599}

\bibitem[\protect\citeauthoryear{{Dale}, {Ercolano}  \& {Bonnell}}{{Dale}
  et~al.}{2012}]{2012MNRAS.424..377D}
{Dale} J.~E.,  {Ercolano} B.,   {Bonnell} I.~A.,  2012, \mn@doi [\mnras]
  {10.1111/j.1365-2966.2012.21205.x}, \href
  {http://esoads.eso.org/abs/2012MNRAS.424..377D} {424, 377}

\bibitem[\protect\citeauthoryear{{Dale}, {Ngoumou}, {Ercolano}  \&
  {Bonnell}}{{Dale} et~al.}{2014}]{2014MNRAS.442..694D}
{Dale} J.~E.,  {Ngoumou} J.,  {Ercolano} B.,   {Bonnell} I.~A.,  2014, \mn@doi
  [\mnras] {10.1093/mnras/stu816}, \href
  {http://esoads.eso.org/abs/2014MNRAS.442..694D} {442, 694}

\bibitem[\protect\citeauthoryear{{Dobbs} \& {Pringle}}{{Dobbs} \&
  {Pringle}}{2013}]{2013MNRAS.432..653D}
{Dobbs} C.~L.,  {Pringle} J.~E.,  2013, \mn@doi [\mnras]
  {10.1093/mnras/stt508}, \href
  {http://adsabs.harvard.edu/abs/2013MNRAS.432..653D} {432, 653}

\bibitem[\protect\citeauthoryear{{Dobbs}, {Bonnell}  \& {Pringle}}{{Dobbs}
  et~al.}{2006}]{2006MNRAS.371.1663D}
{Dobbs} C.~L.,  {Bonnell} I.~A.,   {Pringle} J.~E.,  2006, \mn@doi [\mnras]
  {10.1111/j.1365-2966.2006.10794.x}, \href
  {http://adsabs.harvard.edu/abs/2006MNRAS.371.1663D} {371, 1663}

\bibitem[\protect\citeauthoryear{{Dobbs}, {Pringle}  \& {Burkert}}{{Dobbs}
  et~al.}{2012}]{2012MNRAS.425.2157D}
{Dobbs} C.~L.,  {Pringle} J.~E.,   {Burkert} A.,  2012, \mn@doi [\mnras]
  {10.1111/j.1365-2966.2012.21558.x}, \href
  {http://adsabs.harvard.edu/abs/2012MNRAS.425.2157D} {425, 2157}

\bibitem[\protect\citeauthoryear{{Falceta-Gon{\c c}alves}, {Kowal}, {Falgarone}
   \& {Chian}}{{Falceta-Gon{\c c}alves} et~al.}{2014}]{2014NPGeo..21..587F}
{Falceta-Gon{\c c}alves} D.,  {Kowal} G.,  {Falgarone} E.,   {Chian} A.~C.-L.,
  2014, \mn@doi [Nonlinear Processes in Geophysics] {10.5194/npg-21-587-2014},
  \href {http://adsabs.harvard.edu/abs/2014NPGeo..21..587F} {21, 587}

\bibitem[\protect\citeauthoryear{{Falceta-Gon{\c c}alves}, {Bonnell}, {Kowal},
  {L{\'e}pine}  \& {Braga}}{{Falceta-Gon{\c c}alves}
  et~al.}{2015}]{2015MNRAS.446..973F}
{Falceta-Gon{\c c}alves} D.,  {Bonnell} I.,  {Kowal} G.,  {L{\'e}pine}
  J.~R.~D.,   {Braga} C.~A.~S.,  2015, \mn@doi [\mnras]
  {10.1093/mnras/stu2127}, \href
  {http://adsabs.harvard.edu/abs/2015MNRAS.446..973F} {446, 973}

\bibitem[\protect\citeauthoryear{{Federrath} \& {Klessen}}{{Federrath} \&
  {Klessen}}{2012}]{2012ApJ...761..156F}
{Federrath} C.,  {Klessen} R.~S.,  2012, \mn@doi [\apj]
  {10.1088/0004-637X/761/2/156}, \href
  {http://adsabs.harvard.edu/abs/2012ApJ...761..156F} {761, 156}

\bibitem[\protect\citeauthoryear{{Glover} \& {Clark}}{{Glover} \&
  {Clark}}{2012}]{2012MNRAS.421..116G}
{Glover} S.~C.~O.,  {Clark} P.~C.,  2012, \mn@doi [\mnras]
  {10.1111/j.1365-2966.2011.20260.x}, \href
  {http://adsabs.harvard.edu/abs/2012MNRAS.421..116G} {421, 116}

\bibitem[\protect\citeauthoryear{{Hennebelle} \& {Chabrier}}{{Hennebelle} \&
  {Chabrier}}{2013}]{2013ApJ...770..150H}
{Hennebelle} P.,  {Chabrier} G.,  2013, \mn@doi [\apj]
  {10.1088/0004-637X/770/2/150}, \href
  {http://adsabs.harvard.edu/abs/2013ApJ...770..150H} {770, 150}

\bibitem[\protect\citeauthoryear{{Kennicutt} \& {Evans}}{{Kennicutt} \&
  {Evans}}{2012}]{2012ARA&A..50..531K}
{Kennicutt} R.~C.,  {Evans} N.~J.,  2012, \mn@doi [\araa]
  {10.1146/annurev-astro-081811-125610}, \href
  {http://adsabs.harvard.edu/abs/2012ARA%26A..50..531K} {50, 531}

\bibitem[\protect\citeauthoryear{{Koyama} \& {Inutsuka}}{{Koyama} \&
  {Inutsuka}}{2000}]{2000ApJ...532..980K}
{Koyama} H.,  {Inutsuka} S.-I.,  2000, \mn@doi [\apj] {10.1086/308594}, \href
  {http://adsabs.harvard.edu/abs/2000ApJ...532..980K} {532, 980}

\bibitem[\protect\citeauthoryear{{Koyama} \& {Inutsuka}}{{Koyama} \&
  {Inutsuka}}{2002}]{2002ApJ...564L..97K}
{Koyama} H.,  {Inutsuka} S.-i.,  2002, \mn@doi [\apjl] {10.1086/338978}, \href
  {http://adsabs.harvard.edu/abs/2002ApJ...564L..97K} {564, L97}

\bibitem[\protect\citeauthoryear{{Krumholz} \& {McKee}}{{Krumholz} \&
  {McKee}}{2005}]{2005ApJ...630..250K}
{Krumholz} M.~R.,  {McKee} C.~F.,  2005, \mn@doi [\apj] {10.1086/431734}, \href
  {http://adsabs.harvard.edu/abs/2005ApJ...630..250K} {630, 250}

\bibitem[\protect\citeauthoryear{{Krumholz}, {Klein}  \& {McKee}}{{Krumholz}
  et~al.}{2011}]{2011ApJ...740...74K}
{Krumholz} M.~R.,  {Klein} R.~I.,   {McKee} C.~F.,  2011, \mn@doi [\apj]
  {10.1088/0004-637X/740/2/74}, \href
  {http://adsabs.harvard.edu/abs/2011ApJ...740...74K} {740, 74}

\bibitem[\protect\citeauthoryear{{Larson}, {Evans}, {Green}  \&
  {Yang}}{{Larson} et~al.}{2015}]{2015ApJ...806...70L}
{Larson} R.~L.,  {Evans} II N.~J.,  {Green} J.~D.,   {Yang} Y.-L.,  2015,
  \mn@doi [\apj] {10.1088/0004-637X/806/1/70}, \href
  {http://adsabs.harvard.edu/abs/2015ApJ...806...70L} {806, 70}

\bibitem[\protect\citeauthoryear{{Louvet} et~al.,}{{Louvet}
  et~al.}{2014}]{2014A&A...570A..15L}
{Louvet} F.,  et~al., 2014, \mn@doi [\aap] {10.1051/0004-6361/201423603}, \href
  {http://adsabs.harvard.edu/abs/2014A%26A...570A..15L} {570, A15}

\bibitem[\protect\citeauthoryear{{MacLachlan}, {Bonnell}, {Wood}  \&
  {Dale}}{{MacLachlan} et~al.}{2015}]{2015A&A...573A.112M}
{MacLachlan} J.~M.,  {Bonnell} I.~A.,  {Wood} K.,   {Dale} J.~E.,  2015,
  \mn@doi [\aap] {10.1051/0004-6361/201322250}, \href
  {http://adsabs.harvard.edu/abs/2015A%26A...573A.112M} {573, A112}

\bibitem[\protect\citeauthoryear{{Micic}, {Glover}, {Banerjee}  \&
  {Klessen}}{{Micic} et~al.}{2013}]{2013MNRAS.432..626M}
{Micic} M.,  {Glover} S.~C.~O.,  {Banerjee} R.,   {Klessen} R.~S.,  2013,
  \mn@doi [\mnras] {10.1093/mnras/stt489}, \href
  {http://adsabs.harvard.edu/abs/2013MNRAS.432..626M} {432, 626}

\bibitem[\protect\citeauthoryear{{Nguyen Luong} et~al.,}{{Nguyen Luong}
  et~al.}{2011}]{2011A&A...535A..76N}
{Nguyen Luong} Q.,  et~al., 2011, \mn@doi [\aap] {10.1051/0004-6361/201117831},
  \href {http://adsabs.harvard.edu/abs/2011A%26A...535A..76N} {535, A76}

\bibitem[\protect\citeauthoryear{{Padoan} \& {Nordlund}}{{Padoan} \&
  {Nordlund}}{2011}]{2011ApJ...730...40P}
{Padoan} P.,  {Nordlund} {\AA}.,  2011, \mn@doi [\apj]
  {10.1088/0004-637X/730/1/40}, \href
  {http://adsabs.harvard.edu/abs/2011ApJ...730...40P} {730, 40}

\bibitem[\protect\citeauthoryear{{Palau} et~al.,}{{Palau}
  et~al.}{2013}]{2013ApJ...762..120P}
{Palau} A.,  et~al., 2013, \mn@doi [\apj] {10.1088/0004-637X/762/2/120}, \href
  {http://adsabs.harvard.edu/abs/2013ApJ...762..120P} {762, 120}

\bibitem[\protect\citeauthoryear{{Peretto} et~al.,}{{Peretto}
  et~al.}{2013}]{2013A&A...555A.112P}
{Peretto} N.,  et~al., 2013, \mn@doi [\aap] {10.1051/0004-6361/201321318},
  \href {http://adsabs.harvard.edu/abs/2013A%26A...555A.112P} {555, A112}

\bibitem[\protect\citeauthoryear{{Price} \& {Bate}}{{Price} \&
  {Bate}}{2008}]{2008MNRAS.385.1820P}
{Price} D.~J.,  {Bate} M.~R.,  2008, \mn@doi [\mnras]
  {10.1111/j.1365-2966.2008.12976.x}, \href
  {http://adsabs.harvard.edu/abs/2008MNRAS.385.1820P} {385, 1820}

\bibitem[\protect\citeauthoryear{{Price} \& {Bate}}{{Price} \&
  {Bate}}{2009}]{2009MNRAS.398...33P}
{Price} D.~J.,  {Bate} M.~R.,  2009, \mn@doi [\mnras]
  {10.1111/j.1365-2966.2009.14969.x}, \href
  {http://adsabs.harvard.edu/abs/2009MNRAS.398...33P} {398, 33}

\bibitem[\protect\citeauthoryear{{Smith}, {Glover}, {Clark}, {Klessen}  \&
  {Springel}}{{Smith} et~al.}{2014}]{2014MNRAS.441.1628S}
{Smith} R.~J.,  {Glover} S.~C.~O.,  {Clark} P.~C.,  {Klessen} R.~S.,
  {Springel} V.,  2014, \mn@doi [\mnras] {10.1093/mnras/stu616}, \href
  {http://adsabs.harvard.edu/abs/2014MNRAS.441.1628S} {441, 1628}

\bibitem[\protect\citeauthoryear{{V{\'a}zquez-Semadeni}, {G{\'o}mez},
  {Jappsen}, {Ballesteros-Paredes}, {Gonz{\'a}lez}  \&
  {Klessen}}{{V{\'a}zquez-Semadeni} et~al.}{2007}]{2007ApJ...657..870V}
{V{\'a}zquez-Semadeni} E.,  {G{\'o}mez} G.~C.,  {Jappsen} A.~K.,
  {Ballesteros-Paredes} J.,  {Gonz{\'a}lez} R.~F.,   {Klessen} R.~S.,  2007,
  \mn@doi [\apj] {10.1086/510771}, \href
  {http://adsabs.harvard.edu/abs/2007ApJ...657..870V} {657, 870}

\makeatother
\end{thebibliography}


%\begin{thebibliography}{99}
%\end{thebibliography}

%%%%%%%%%%%%%%%%%%%%%%%%%%%%%%%%%%%%%%%%%%%%%%%%%%

%%%%%%%%%%%%%%%%% APPENDICES %%%%%%%%%%%%%%%%%%%%%

%\appendix



%%%%%%%%%%%%%%%%%%%%%%%%%%%%%%%%%%%%%%%%%%%%%%%%%%

% Don't change these lines
\bsp	% typesetting comment
\label{lastpage}
\end{document}